\documentclass[aps,pre,twocolumn,superscriptaddress]{revtex4-2}

\usepackage{amsmath}
\usepackage{graphicx}
\usepackage{dcolumn}
\usepackage{bm}
\usepackage{amssymb}
\usepackage{xcolor}
\usepackage[colorlinks=true, linkcolor=blue, citecolor=blue, urlcolor=blue]{hyperref}
\usepackage{orcidlink} 

\begin{document}

\preprint{APS/123-QED}

\title{Sorted Mutability as an Entropic Functional: Theory and Two Applications}

\author{Michel Aguilera \orcidlink{0000-0002-3049-668X}}
\email{michel.aguilera@uss.cl} 
\affiliation{Instituto de Física, Pontificia Universidad Católica de Valparaíso, Casilla 4950, Valparaíso 2373223, Chile}
\affiliation{Departamento de Física, Universidad Técnica Federico Santa María, Av. España 1680, Valparaíso 2390123, Chile}

\author{Eugenio E. Vogel \orcidlink{0000-0002-1701-3030}}
\affiliation{Departamento de Ciencias Físicas, Universidad de La Frontera, Avenida Francisco Salazar 01145, Temuco, Chile}
\affiliation{CEDENNA, FINARQ, Universidad Central de Chile, Av. Santa Isabel 1186, Santiago 8330601, Chile}

\author{Francisco J. Pe\~na \orcidlink{0000-0002-7432-0707}}
\affiliation{Facultad de Ingeniería, Universidad San Sebastián,
Lago Panguipulli 1390, 5501842 Puerto Montt, Chile}

\author{Gonzalo Saravia}
\affiliation{Los Eucaliptus 1189, Temuco 4812537, Chile}

\author{Patricio Vargas \orcidlink{0000-0001-9235-9747}}
\affiliation{Departamento de Física, Universidad Técnica Federico Santa María, Av. España 1680, Valparaíso 2390123, Chile}

\date{\today}

\begin{abstract}
We investigate sorted mutability, $\zeta_S$, as an information-based quantity with an explicit entropic character. We show that $\zeta_S$ is directly proportional to the diversity of a data sequence and can therefore be expressed in terms of Shannon entropies, establishing a connection between compression-based measures and conventional information theory. In contrast to regular mutability, which preserves the sequential ordering of the data, sorted mutability removes this ordering and provides a static characterization of the distribution of observed values. We illustrate these properties in two complementary settings: the two-dimensional ferromagnetic $Q$-clock model and wind-power production data. For the $Q$-clock model, sorted mutability and diversity display the same temperature dependence, consistent with their analytical relation, while conventional thermodynamic observables provide an independent reference. In the case of wind power, we report monthly results for sorted mutability and diversity, showing proportional time evolutions. These results support sorted mutability as a useful, easy-to-calculate entropic functional for analyzing both equilibrium and non-equilibrium data sets.
\end{abstract}
\maketitle
\section{Introduction}
\label{introduccion}

Entropy provides one of the most fundamental links between microscopic description, macroscopic behavior, and information. In equilibrium statistical mechanics, the ideas introduced by Boltzmann and Gibbs established entropy as a measure associated with the statistical weight of microscopic configurations \cite{boltzmann1877beziehung,gibbs1902elementary,wehrl1978general}. The information-theoretic formulation of Shannon, followed by the maximum-entropy approach of Jaynes, further clarified the connection between statistical mechanics and inference, establishing a bridge that remains central in modern statistical physics \cite{shannon1948mathematical,cover2006elements,jaynes1957information,landauer1961irreversibility,parrondo2015thermodynamics}

The success of this framework stimulated the development of generalized entropic measures, such as Rényi and Tsallis entropies, intended for systems whose statistical organization requires more than a single conventional entropy \cite{renyi1961measures,tsallis1988possible,abe2000axioms,beck2003superstatistics,hanel2011classification,tempesta2016beyond,balogh2020generalized}. These approaches emphasize that an entropic functional is not merely a numerical measure of disorder, but a mapping that reflects specific statistical constraints and structures.

When analyzing finite experimental or numerical sequences, information resides both in the abundances of observed values and in their sequential organization. Algorithmic information theory and practical compression algorithms highlighted this distinction early on \cite{kolmogorov1965three,chaitin1966length,lempel1976complexity,ziv1977universal}. Consequently, various complexity metrics have been developed to characterize time series, demonstrating that different information measures encode distinct properties of a data sequence \cite{crutchfield1989inferring,bandt2002permutation,costa2002multiscale,rosso2007distinguishing}.

A complementary descriptor is diversity, which quantifies the richness of the set of distinguishable values represented in a sample \cite{hill1973diversity,jost2006entropy,leinster2012measuring,chao2014unifying}. Recently, diversity defined as the fraction of different outcomes was shown to admit an explicit representation as a difference of Shannon entropies \cite{vogel2026diversity}. Within this general framework, \emph{mutability} was introduced as an information-based quantity designed to characterize sequences through lossless compression \cite{vogel2009phasetransitions,vogel2012data}. Unlike Shannon entropy, regular mutability is sensitive to chronological ordering and has been successfully applied to data sets of very different origins \cite{vogel2014EPJB,contreras2016bloodpressure,vogel2017rods,vogel2018wind,vogel2020measuring,pasten2022information}.

A related quantity, \emph{sorted mutability}, is obtained by applying the compression procedure after rearranging the original data in increasing or decreasing order \cite{vogel2025entropy}. Sorting deliberately removes the original sequential organization while preserving the represented values and their abundances, isolating a static property of the set. 

The purpose of the present work is to establish the analytical meaning of this order-independent quantity. We show that, under a fixed numerical representation and encoding prescription, sorted mutability $\zeta_S$ is directly proportional to the diversity $\delta$ of the sequence. Because diversity can be expressed in terms of Shannon entropy \cite{vogel2026diversity}, sorted mutability inherits an explicit entropic representation, thereby analytically linking a compression-based operational definition to conventional information theory.

We examine this relation in two physically very different settings. The first is the two-dimensional ferromagnetic $Q$-state clock model, a paradigmatic system interpolating between discrete spin models and the continuous planar-rotor limit. Its phase structure, driven by the competition between discrete symmetry and vortex excitations, provides a controlled equilibrium benchmark to compare sorted mutability against Shannon entropy and conventional thermodynamic observables \cite{berezinskii1971destruction,kosterlitz1973ordering,jose1977renormalization,elitzur1979phase,lapilli2006universality,kumano2013response,li2020critical,negrete2018entropy,negrete2021bkt}.

The second application concerns wind-power generation in Germany \cite{vogel2018wind,vogel2024onshore}. Unlike equilibrium Monte Carlo sequences, these records arise from a real, externally driven non-equilibrium process. By evaluating the same analytical relation in both a controlled equilibrium model and an empirical non-equilibrium data set, our objective is to clarify the exact entropic interpretation of sorted mutability and to identify explicitly the static information it retains when sequence ordering is eliminated.

\section{Theory and definitions}
\label{sec:theory}

Let us consider a finite experimental, empirical, or numerical data sequence containing $\rho$ measurements of the same observable, $\mathcal{W}=\{W_1,W_2,\ldots,W_i,\ldots,W_{\rho}\}$. All records are assumed to have been obtained using the same measurement or numerical prescription and are therefore assigned equal statistical weight independently of their measured value. We thus introduce the uniform probability distribution
\begin{equation}
p_i=\frac{1}{\rho},\qquad i=1,2,\ldots,\rho.
\label{eq:pi_measurements}
\end{equation}

From the original sequence we can construct a second sequence containing only the $\rho^\ast$ distinct values represented in $\mathcal{W}$, where by construction $\rho^\ast\leq\rho$. Since only the distinction between different represented values is relevant in this auxiliary sequence, each of its entries is assigned the uniform probability
\begin{equation}
p_i^\ast=\frac{1}{\rho^\ast},\qquad i=1,2,\ldots,\rho^\ast.
\label{eq:pi_distinct}
\end{equation}
When convenient, both distributions may be embedded in a common $\rho$-dimensional representation by defining $p_i^\ast=0$ for $\rho^\ast<i\leq\rho$, with the usual convention $0\ln0=0$.

A third probability distribution accounts for the actual abundances of the distinct measured values. If the $i$th distinct value appears $f_i$ times in the original sequence, with
\begin{equation}
\sum_{i=1}^{\rho^\ast}f_i=\rho,
\label{eq:frequency_sum}
\end{equation}
its empirical probability is
\begin{equation}
P_i=\frac{f_i}{\rho},\qquad i=1,2,\ldots,\rho^\ast.
\label{eq:Pi_frequency}
\end{equation}
The distinction between $p_i$, $p_i^\ast$, and $P_i$ will be important below: the first two distributions are uniform constructions associated with the total and distinct records, respectively, whereas $P_i$ contains the empirical abundance information of the measured values.

To make these definitions explicit, Table~\ref{tab:wind} presents a short illustrative sequence based on wind-power generation in Germany on January 28, 2023. The sequence contains $\rho=32$ records, corresponding to measurements taken every 15 minutes over an eight-hour interval. Column 1 gives the record index, while column 2 lists the generated power in GW. A few numerical values are slightly adjusted for illustrative purposes so that repetitions and the construction of sorted mutability can be displayed transparently; this example is not used in the quantitative wind-power analysis presented later.

Column 3 gives the same data in TW using a fixed numerical representation consisting of two integer digits and one decimal digit, with leading zeros where required. This fixed-width representation is essential because the compression-based quantities introduced below depend on the number of bytes used to encode each record. Column 4 contains the same sequence sorted in increasing order. Column 5 lists each distinct represented value together with its frequency $f_i$. For this example there are $\rho^\ast=18$ distinct values among the $\rho=32$ records.

\begin{table}[t]
\centering
\caption{Illustrative sequence of wind-power records sampled every 15 minutes over an eight-hour interval on January 28, 2023. Column 1 gives the record index $i$; column 2 gives the generated power in Gw; column 3 gives the same power in Tw using the fixed numerical representation adopted for the information-based analysis; column 4 gives the sequence sorted in increasing order; and column 5 lists the $\rho^\ast$ distinct represented values and their frequencies $f_i$.}
\label{tab:wind}
\small
\setlength{\tabcolsep}{4pt}
\begin{tabular}{|r|r|c|c|c|}
\hline
$i$ & Gw & $\mathbf{Tw}$ & $S$ & Distinct value $\;\; f_i$ \\
\hline
1  & 7641  & $\mathbf{07.6}$ & 07.6 & 07.6 $\;\;$ 01 \\
2  & 8003  & $\mathbf{08.0}$ & 08.0 & 08.0 $\;\;$ 01 \\
3  & 8485  & $\mathbf{08.5}$ & 08.5 & 08.5 $\;\;$ 01 \\
4  & 8857  & $\mathbf{08.9}$ & 08.9 & 08.9 $\;\;$ 01 \\
5  & 9177  & $\mathbf{09.2}$ & 09.2 & 09.2 $\;\;$ 02 \\
6  & 9242  & $\mathbf{09.2}$ & 09.2 & 09.5 $\;\;$ 02 \\
7  & 9482  & $\mathbf{09.5}$ & 09.5 & 09.6 $\;\;$ 01 \\
8  & 9588  & $\mathbf{09.6}$ & 09.5 & 09.9 $\;\;$ 01 \\
\hline
9  & 9512  & $\mathbf{09.5}$ & 09.6 & 10.0 $\;\;$ 01 \\
10 & 9928  & $\mathbf{09.9}$ & 09.9 & 10.1 $\;\;$ 04 \\
11 & 10015 & $\mathbf{10.0}$ & 10.0 & 10.2 $\;\;$ 08 \\
12 & 10071 & $\mathbf{10.1}$ & 10.1 & 10.3 $\;\;$ 02 \\
13 & 10071 & $\mathbf{10.1}$ & 10.1 & 10.5 $\;\;$ 01 \\
14 & 10209 & $\mathbf{10.2}$ & 10.1 & 10.8 $\;\;$ 01 \\
15 & 10159 & $\mathbf{10.2}$ & 10.1 & 11.3 $\;\;$ 01 \\
16 & 10270 & $\mathbf{10.3}$ & 10.2 & 11.9 $\;\;$ 02 \\
\hline
17 & 10240 & $\mathbf{10.2}$ & 10.2 & 12.1 $\;\;$ 01 \\
18 & 10188 & $\mathbf{10.2}$ & 10.2 & 12.4 $\;\;$ 01 \\
19 & 10154 & $\mathbf{10.2}$ & 10.2 & \\
20 & 10143 & $\mathbf{10.1}$ & 10.2 & \\
21 & 10177 & $\mathbf{10.2}$ & 10.2 & \\
22 & 10090 & $\mathbf{10.1}$ & 10.2 & \\
23 & 10230 & $\mathbf{10.2}$ & 10.2 & \\
24 & 10211 & $\mathbf{10.2}$ & 10.3 & \\
\hline
25 & 10273 & $\mathbf{10.3}$ & 10.3 & \\
26 & 10466 & $\mathbf{10.5}$ & 10.5 & \\
27 & 10830 & $\mathbf{10.8}$ & 10.8 & \\
28 & 11259 & $\mathbf{11.3}$ & 11.3 & \\
29 & 11861 & $\mathbf{11.9}$ & 11.9 & \\
30 & 11883 & $\mathbf{11.9}$ & 11.9 & \\
31 & 12125 & $\mathbf{12.1}$ & 12.1 & \\
32 & 12395 & $\mathbf{12.4}$ & 12.4 & \\
\hline
\end{tabular}
\end{table}

We now introduce the three information quantities required for the subsequent analysis, using Table~\ref{tab:wind} as an explicit example.

\noindent\textbf{Shannon entropies.} For a discrete probability distribution $\{q_i\}$, the Shannon entropy is defined as $-\sum_i q_i\ln q_i$ \cite{shannon1948mathematical}. Applying this definition to the empirical abundance distribution $P_i$ gives
\begin{equation}
H=-\sum_{i=1}^{\rho^\ast}P_i\ln P_i.
\label{eq:shannon_empirical}
\end{equation}
The corresponding entropies associated with the two uniform distributions defined in Eqs.~(\ref{eq:pi_measurements}) and (\ref{eq:pi_distinct}) are
\begin{equation}
h_M=-\sum_{i=1}^{\rho}p_i\ln p_i=\ln\rho,
\label{eq:hM}
\end{equation}
and
\begin{equation}
h_D=-\sum_{i=1}^{\rho^\ast}p_i^\ast\ln p_i^\ast=\ln\rho^\ast.
\label{eq:hD}
\end{equation}
These entropies have distinct meanings. The empirical Shannon entropy $H$ depends on the full abundance distribution of the represented values, whereas $h_M$ and $h_D$ depend only on the number of total and distinct records, respectively. It is the difference between $h_D$ and $h_M$, rather than the empirical entropy $H$, that enters the exact entropic representation of diversity discussed below.

\noindent\textbf{Diversity.} Diversity is a dimensionless quantity defined as the fraction of distinct represented values in a sequence \cite{vogel2014EPJB},
\begin{equation}
\delta=\frac{\rho^\ast}{\rho}.
\label{eq:diversity}
\end{equation}
For the illustrative sequence in Table~\ref{tab:wind}, $\rho^\ast=18$ and $\rho=32$, giving $\delta=18/32=0.5625$. More recently, diversity was shown to admit an explicit representation in terms of entropic functionals \cite{vogel2026diversity}. Using Eqs.~(\ref{eq:hM}) and (\ref{eq:hD}),
\begin{equation}
h_D-h_M=\ln\rho^\ast-\ln\rho=\ln\left(\frac{\rho^\ast}{\rho}\right),
\end{equation}
and therefore
\begin{equation}
\delta=e^{h_D-h_M}.
\label{eq:diversity_entropy}
\end{equation}
Equation~(\ref{eq:diversity_entropy}) establishes the explicit connection between diversity and the two Shannon entropies associated with the total and distinct-record distributions \cite{vogel2026diversity}.

\noindent\textbf{Sorted mutability.} Sorted mutability is obtained after sorting the original data sequence while retaining a fixed numerical representation and encoding prescription \cite{vogel2025entropy}. In the illustrative example of Table~\ref{tab:wind}, column 4 contains the $\rho$ sorted records. If each encoded record occupies $a$ bytes, the total size of the sorted sequence is
\begin{equation}
w=a\rho.
\label{eq:w_original}
\end{equation}
The same value of $w$ is obtained whether the sequence is sorted in increasing or decreasing order.

The compressed representation of the sorted sequence can be written in terms of its $\rho^\ast$ distinct values and their corresponding abundances. Under the fixed-width encoding adopted here, each distinct-value/frequency record occupies $b$ bytes, so that the size of this representation is
\begin{equation}
w^\dagger=b\rho^\ast.
\label{eq:w_compressed}
\end{equation}
The quantities $a$ and $b$ are therefore determined by the adopted numerical representation and encoding prescription and must remain fixed within a given comparison.

Sorted mutability is defined as the ratio between these two file sizes \cite{vogel2025entropy},
\begin{equation}
\zeta_S=\frac{w^\dagger}{w}=\frac{b\rho^\ast}{a\rho}.
\label{eq:sorted_mutability}
\end{equation}
Using Eq.~(\ref{eq:diversity}), it follows immediately that
\begin{equation}
\zeta_S=\frac{b}{a}\,\delta.
\label{zeta_s_delta}
\end{equation}
Thus, for a fixed numerical representation and encoding prescription, sorted mutability is exactly proportional to diversity. The proportionality factor $b/a$ is representation dependent but independent of the particular ordering of the original sequence. Consequently, all variations of $\zeta_S$ within a comparison performed with fixed $a$ and $b$ originate from the diversity $\rho^\ast/\rho$ of the represented data.

Finally, combining Eq.~(\ref{zeta_s_delta}) with the entropic representation of diversity in Eq.~(\ref{eq:diversity_entropy}) gives
\begin{equation}
\zeta_S=\frac{b}{a}\,e^{h_D-h_M}.
\label{eq:zeta_entropy}
\end{equation}
Sorted mutability can therefore be expressed explicitly in terms of Shannon entropies. This result establishes the central theoretical connection of the present work: a quantity originally defined operationally through a compression procedure reduces, after sorting and under a fixed encoding prescription, to an entropic functional proportional to the diversity of the represented sequence.

\section{Applications}
\label{sec:applications}

We now apply the information-based quantities introduced in Sec.~\ref{sec:theory} to two complementary classes of data. The first consists of energy sequences generated numerically for the two-dimensional ferromagnetic $Q$-state clock model. This system provides a controlled equilibrium setting in which Shannon entropy, diversity, and sorted mutability can be examined against conventional thermodynamic observables. The thermal behavior of the $Q$-clock model depends strongly on the number of allowed spin orientations. In particular, for $Q>4$, the low-temperature ordered phase and the high-temperature paramagnetic phase are separated by an intermediate critical regime associated with vortex excitations, leading to two characteristic thermal transitions \cite{negrete2018entropy,negrete2021bkt}. The second application concerns empirical wind-power generation data from Germany during 2023 and 2024, recorded at 15-minute intervals \cite{German_wind,vogel2024onshore}. Unlike the equilibrium Monte Carlo sequences, these records arise from a real, externally driven energy-generation process. Taken together, these two applications allow us to examine the same information-based relation in simulated equilibrium data and in empirical non-equilibrium sequences of a completely different physical origin.

\subsection{The $Q$-clock model}
\label{sec:qclock}

We consider the ferromagnetic $Q$-state clock model on a two-dimensional square lattice of linear size $L$, containing $N=L^{2}$ sites, with open boundary conditions. At each lattice site $i$, the spin is represented by a planar unit vector restricted to one of $Q$ discrete orientations,
\begin{equation}
\mathbf{S}_{i}=\left(\cos\theta_i,\sin\theta_i\right),\qquad \theta_i=\frac{2\pi n_i}{Q},\qquad n_i=0,1,\ldots,Q-1.
\label{eq:qclock_spin}
\end{equation}
The minimum angular separation between consecutive allowed orientations is therefore
\begin{equation}
\Delta\theta=\frac{2\pi}{Q}.
\label{eq:qclock_delta_theta}
\end{equation}

\begin{figure}
\centering
\includegraphics[width=1.0\columnwidth]{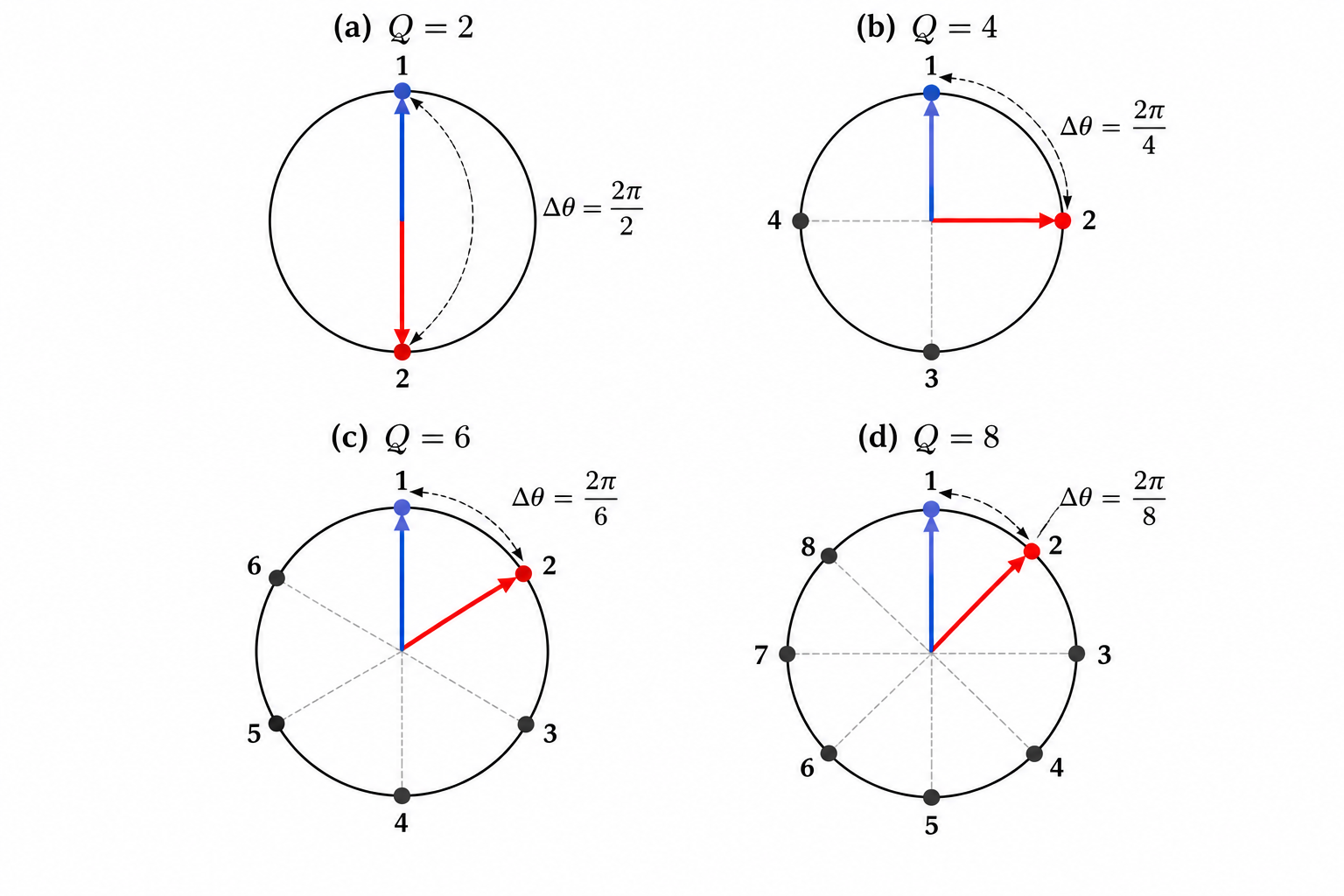}
\caption{Graphical representation of the discrete spin orientations allowed in the $Q$-clock model for $Q=2$, $4$, $6$, and $8$. Each panel displays the $Q$ allowed orientations and highlights two consecutive states separated by the minimum angular interval $\Delta\theta=2\pi/Q$. Odd values of $Q$ are also permitted by the model and were processed but are not shown for space reasons.}
\label{fig:qclock_orientations}
\end{figure}

Figure~\ref{fig:qclock_orientations} illustrates the discrete orientation space for the representative cases $Q=2$, $4$, $6$, and $8$. For $Q=2$, the model is equivalent to the ferromagnetic Ising model, whereas increasing $Q$ progressively enriches the local configuration space and approaches the continuous planar-rotor, or $XY$, limit. Calculations were performed for integer values of $Q$ between 2 and 8; in the main text we present the representative even cases $Q=2$, $4$, and $6$, while an additional result for $Q=8$ is provided in Appendix~B.

The energy of a spin configuration is described by the nearest-neighbor Hamiltonian
\begin{equation}
\mathcal{H}=-J\sum_{\langle i,j\rangle}\cos\left(\theta_i-\theta_j\right),
\label{eq:qclock_hamiltonian}
\end{equation}
where $J>0$ is the ferromagnetic exchange coupling and the sum runs over nearest-neighbor pairs of the square lattice. Throughout the calculations, energy is measured in units of $J$ and temperature in units of $J/k_{\mathrm B}$; equivalently, we set $J=k_{\mathrm B}=1$. No external magnetic field is applied.

The equilibrium configurations are generated using a standard single-spin Metropolis algorithm. At each elementary update, a lattice site $i$ is selected at random and a trial orientation $\theta_i^{\prime}$ is chosen from the set of allowed clock states. The corresponding local energy change is
\begin{equation}
\Delta E=-J\sum_{j\in{\rm nn}(i)}\left[\cos\left(\theta_i^{\prime}-\theta_j\right)-\cos\left(\theta_i-\theta_j\right)\right],
\label{eq:qclock_deltaE}
\end{equation}
where the sum extends over the nearest neighbors of site $i$. The trial state is accepted whenever $\Delta E\leq0$. For $\Delta E>0$, it is accepted with probability
\begin{equation}
P_{\mathrm{acc}}=\exp\left(-\frac{\Delta E}{T}\right).
\label{eq:qclock_metropolis}
\end{equation}
Operationally, the move is accepted when a uniformly distributed random number $r\in[0,1]$ satisfies $r<P_{\mathrm{acc}}$. One Monte Carlo step (MCS) corresponds to $N=L^{2}$ attempted single-spin updates.

For most of the $Q$-clock results presented below, the simulations comprise $1.2\times10^{5}$ MCS, of which the first $6.0\times10^{4}$ MCS are discarded for thermalization and the subsequent trajectory provides $\rho=6.0\times10^{4}$ energy records for the statistical and information-based analysis. In selected calculations, longer sequences containing $\rho=1.2\times10^{5}$ and $\rho=2.4\times10^{5}$ retained records are also considered in order to examine the stability of the information-based quantities with respect to the sequence length. Temperatures are explored sequentially using a cooling protocol from high to low temperature.


After thermalization, the energy is sampled along the Monte Carlo trajectory. A sampled configuration has instantaneous total energy
\begin{equation}
E=-J\sum_{\langle i,j\rangle}\cos\left(\theta_i-\theta_j\right),
\label{eq:qclock_energy}
\end{equation}
while the corresponding energy per spin is
\begin{equation}
u=\frac{E}{N}.
\label{eq:qclock_energy_per_spin}
\end{equation}
The quantity $u$ is used below as a conventional thermodynamic reference. The information-based analysis, however, is performed on the recorded energy sequence
\begin{equation}
\mathcal{E}(T)=\left\{E_1,E_2,\ldots,E_{\rho}\right\},
\label{eq:qclock_energy_sequence}
\end{equation}
where $\rho$ denotes the number of retained energy records at a given temperature.

Because diversity and sorted mutability depend explicitly on the identification of distinct represented values, the numerical representation of the energy records must remain fixed throughout each comparison. Accordingly, the same numerical precision, rounding convention, and encoding prescription are used for all energy sequences entering a given analysis. This requirement is essential because the number of distinct represented energies, $\rho^\ast$, and therefore both $\delta$ and $\zeta_S$, depend on the adopted numerical representation.


From each sequence $\mathcal{E}(T)$ we calculate the empirical Shannon entropy $H$, the diversity $\delta$, and the sorted mutability $\zeta_S$ according to the definitions introduced in Sec.~\ref{sec:theory}. Sorting removes the sequential ordering of the records and therefore eliminates information concerning the temporal arrangement of the sampled energies. Under the fixed encoding prescription, $\zeta_S$ depends on the fraction of distinct represented energy values, $\rho^\ast/\rho$, rather than on their order of occurrence. This property is precisely what distinguishes sorted mutability from order-sensitive information measures and underlies its relation to diversity established in Sec.~\ref{sec:theory}.

\begin{figure}
\centering
\includegraphics[width=0.99\columnwidth]{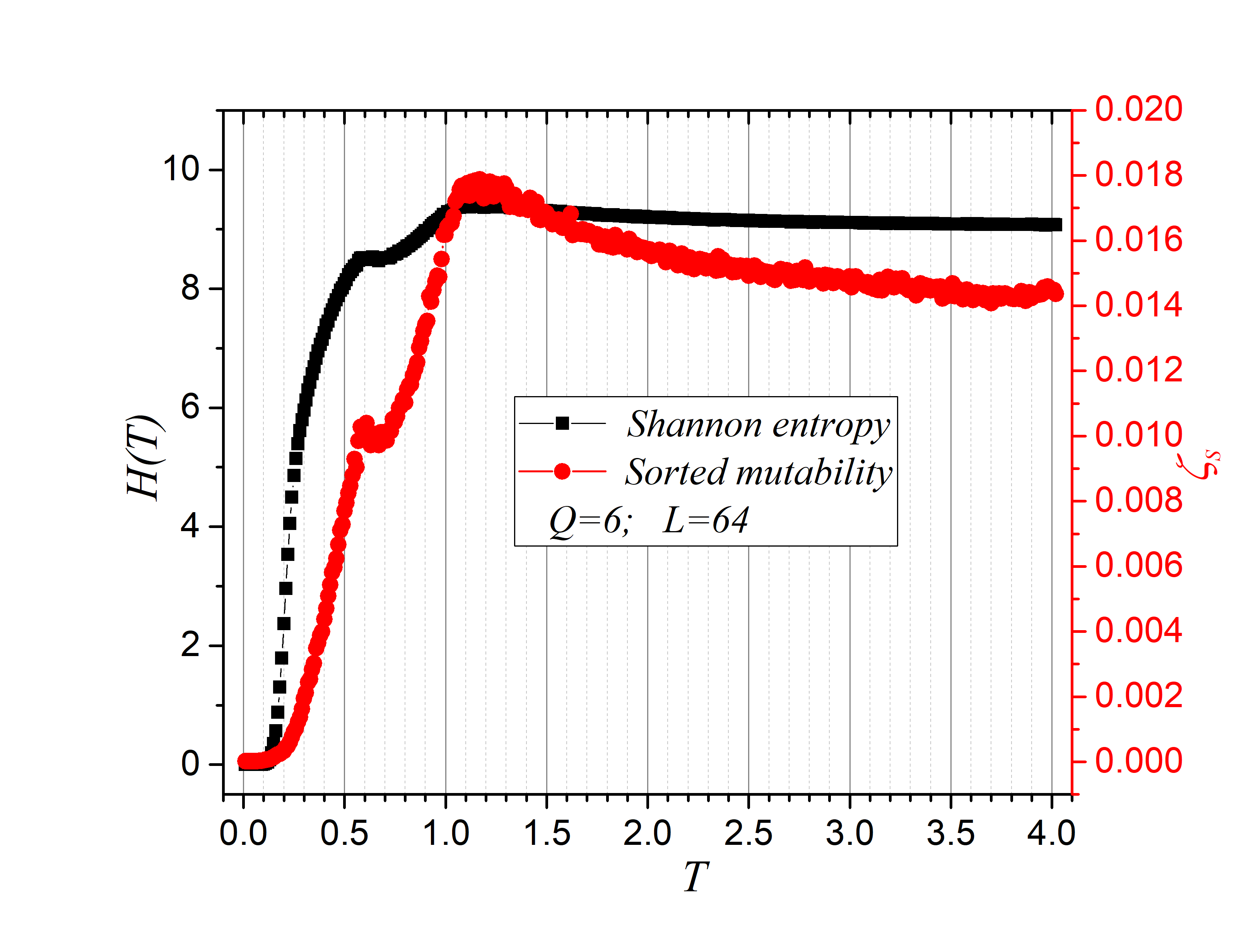}
\caption{Shannon entropy, $H$ (black squares), and sorted mutability, $\zeta_S$ (red circles), as functions of temperature for the $Q=6$ clock model on an $L=64$ square lattice. Both quantities are evaluated from the Monte Carlo energy sequences using the same numerical representation at every temperature. Although their numerical values and functional forms differ, their dominant variations occur over comparable temperature intervals.}
\label{fig:qclock_shannon_sorted}
\end{figure}

We first compare sorted mutability with the empirical Shannon entropy for the representative case $Q=6$ and $L=64$. Figure~\ref{fig:qclock_shannon_sorted} shows $H(T)$ and $\zeta_S(T)$. The two quantities probe different properties of the sampled energy distribution and therefore should not be expected to coincide quantitatively or to have identical functional forms. Nevertheless, they display a clear qualitative co-variation with temperature: their most pronounced changes occur over comparable thermal intervals, and their dominant structures are located within the same broad temperature region. Shannon entropy evolves more smoothly and exhibits broader features, whereas sorted mutability produces a sharper response. The comparison therefore indicates that both quantities are sensitive to changes in the statistical organization of the sampled energy landscape, although they encode different aspects of that distribution.

This distinction is important. The empirical Shannon entropy $H$ depends on the complete frequency distribution of the represented energy values, whereas sorted mutability is linked, under the fixed encoding prescription, to the fraction of distinct values present in the sequence. Similar thermal trends should therefore be interpreted as a qualitative correspondence between two distinct information-based descriptions, rather than as evidence of their identity.

\begin{figure}
\centering
\includegraphics[width=0.99\columnwidth]{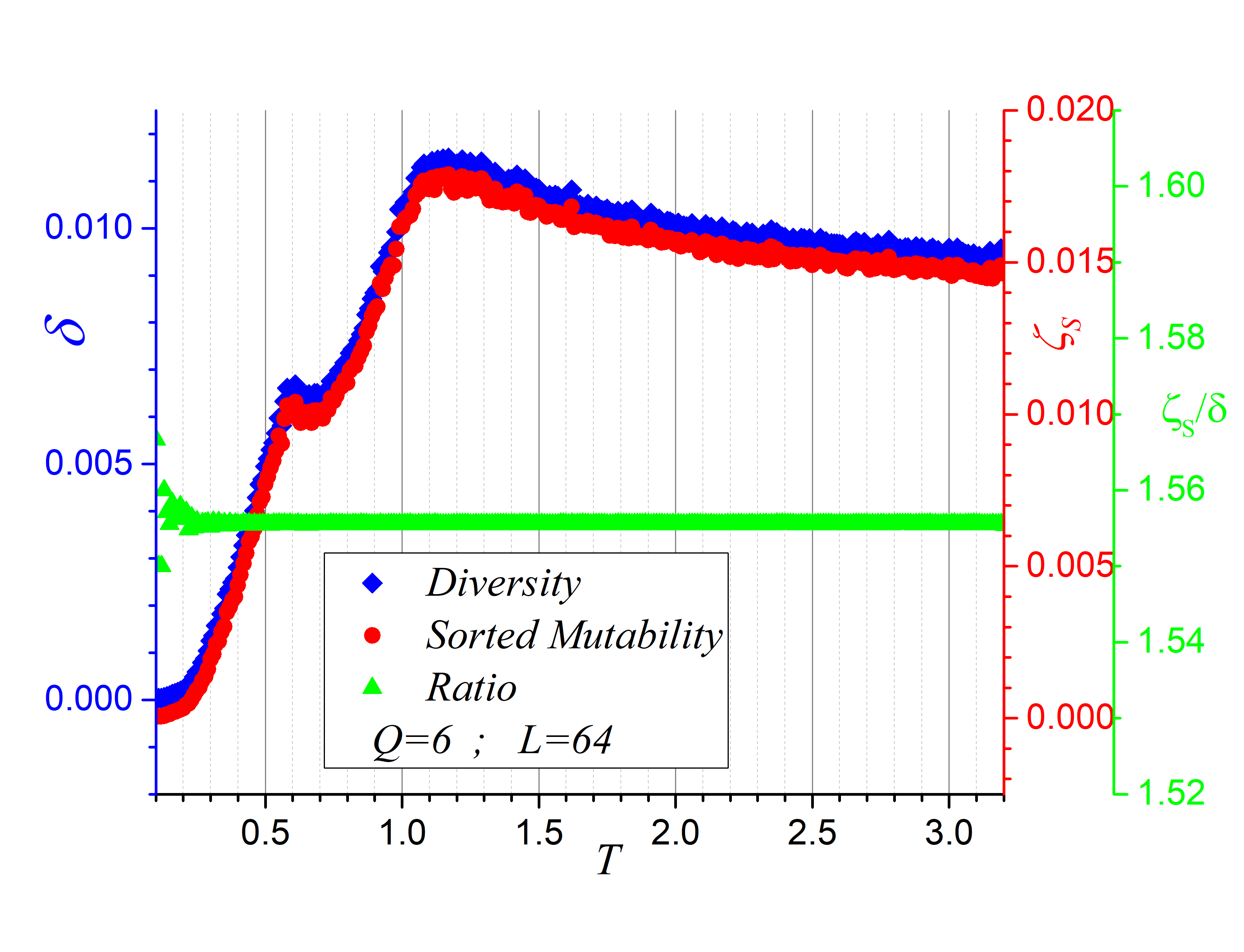}
\caption{Diversity, $\delta$ (blue diamonds), sorted mutability, $\zeta_S$ (red circles), and their ratio $\zeta_S/\delta$ (green triangles), as functions of temperature for the $Q=6$ clock model with $L=64$. The close correspondence between $\delta$ and $\zeta_S$ illustrates the proportionality $\zeta_S=(b/a)\delta$ for the adopted numerical representation.}
\label{fig:qclock_ratio}
\end{figure}

A more direct consistency test follows from the analytical relation between sorted mutability and diversity derived in Eq.~(\ref{zeta_s_delta}). For a fixed representation, $\zeta_S(T)$ and $\delta(T)$ are therefore expected to have the same temperature dependence, differing only by the representation-dependent factor $b/a$. Figure~\ref{fig:qclock_ratio} illustrates this relation for $Q=6$ and $L=64$. Diversity and sorted mutability closely track one another over the complete temperature interval, while the ratio $\zeta_S/\delta$ remains approximately temperature independent. The Monte Carlo results therefore provide a numerical consistency check of the proportionality established analytically in Sec.~\ref{sec:theory}.

\begin{figure}
\centering
\includegraphics[width=0.99\columnwidth]{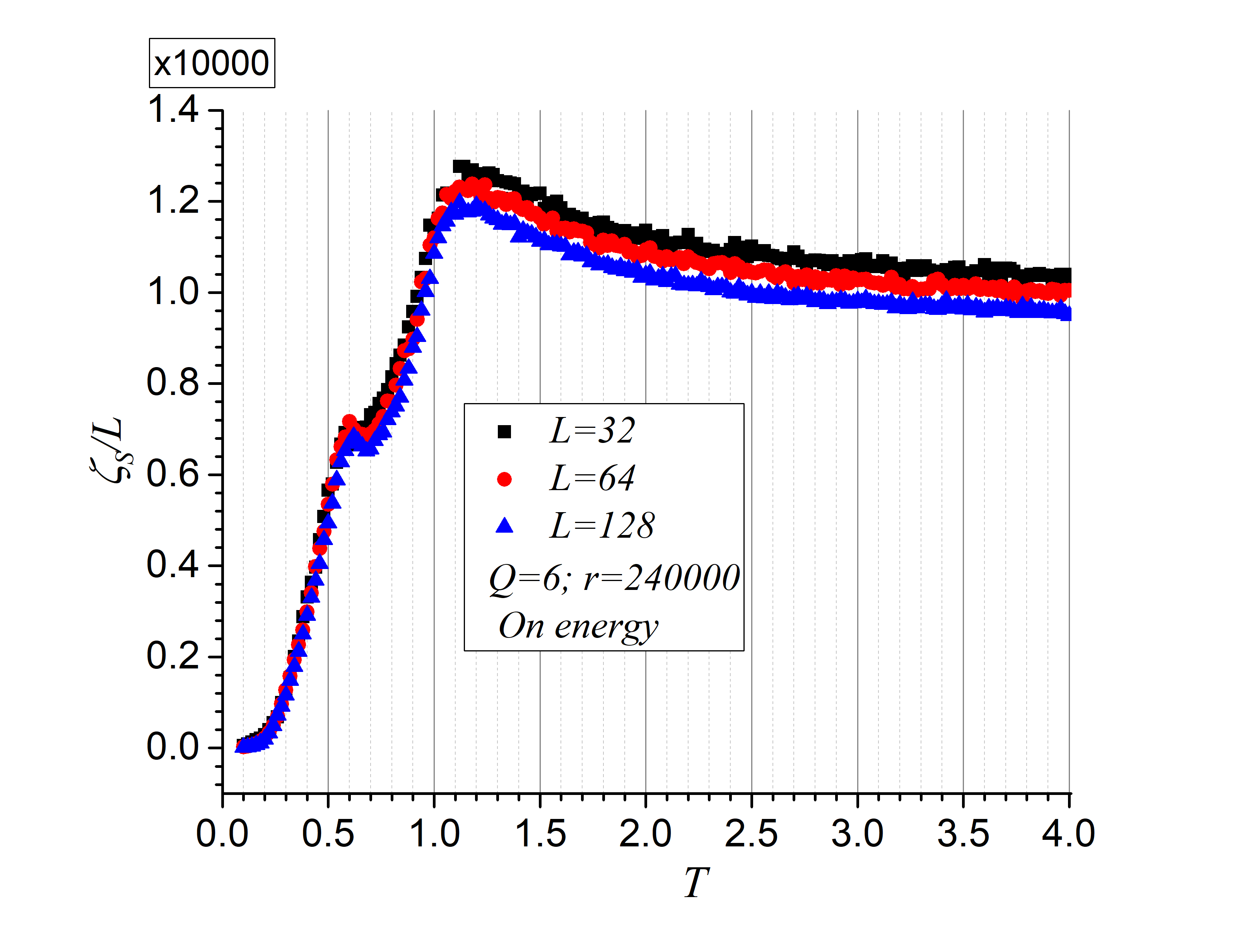}
\caption{Sorted mutability, $\zeta_S$, as a function of temperature for the $Q=6$ clock model with lattice sizes $L=32$, $64$, and $128$, as indicated in the legend. The comparison is intended as a robustness test of the characteristic temperature dependence and not as a finite-size scaling analysis.}
\label{fig:qclock_finite_size}
\end{figure}

We next examine the dependence of the sorted-mutability response on lattice size. Figure~\ref{fig:qclock_finite_size} compares the results obtained for $Q=6$ using $L=32$, $64$, and $128$. Over the range of sizes considered, the curves display the same broad thermal evolution and retain their principal temperature-dependent structures.

This comparison is not intended as a finite-size scaling analysis, and no thermodynamic-limit extrapolation is attempted. Its purpose is instead to determine whether the characteristic information-based structures observed in the Monte Carlo sequences persist as the number of lattice sites is increased. Within the system sizes considered, the principal structures occur in comparable temperature regions. These regions are also relevant to the two-step thermal behavior expected for the $Q=6$ clock model; however, we do not rely solely on the extrema of $\zeta_S$ to identify critical temperatures. Their thermodynamic context is assessed independently using the energy and specific heat presented in Appendix~\ref{app:qclock_thermodynamics}.


\begin{figure}
\centering
\includegraphics[width=0.69\columnwidth]{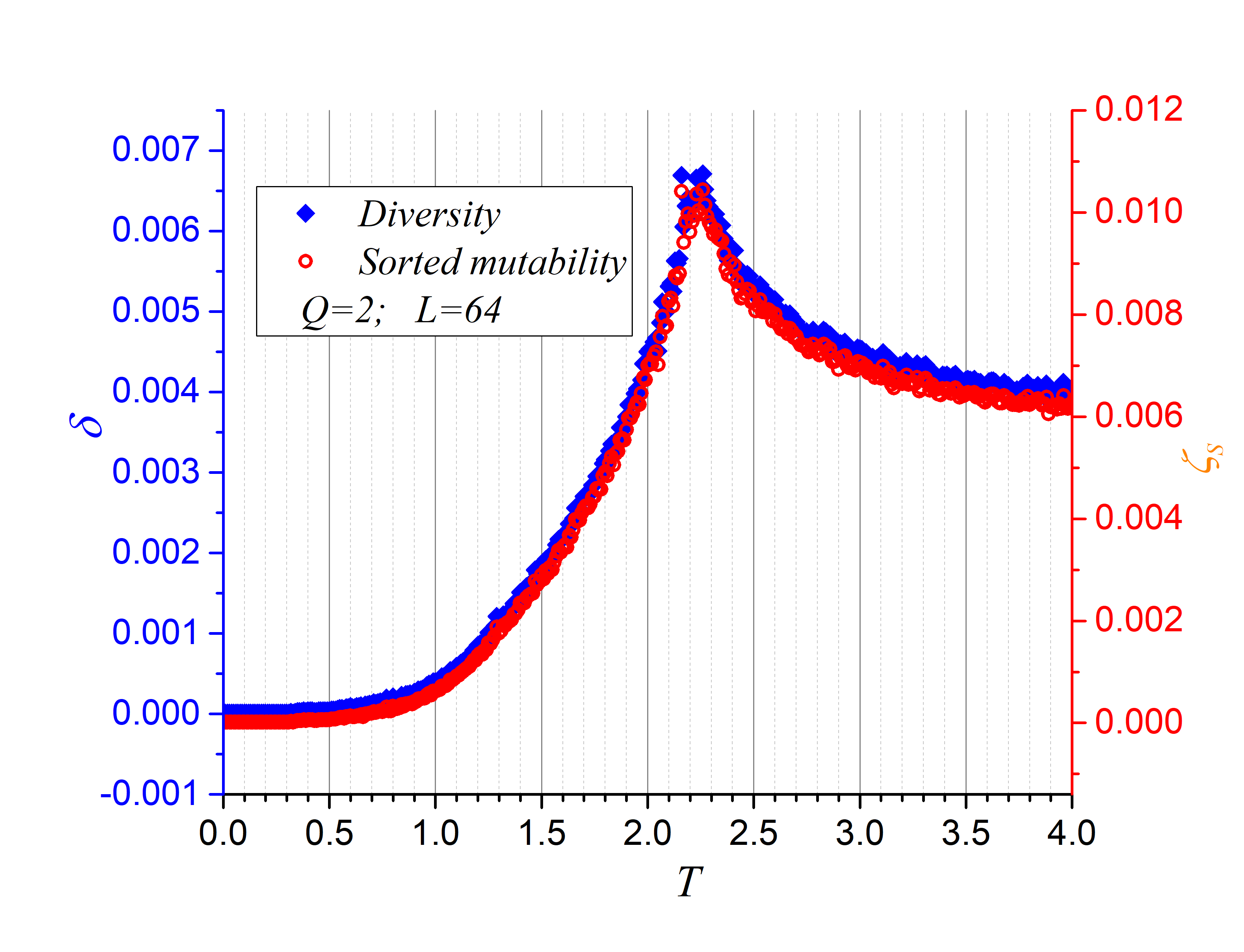}

\includegraphics[width=0.69\columnwidth]{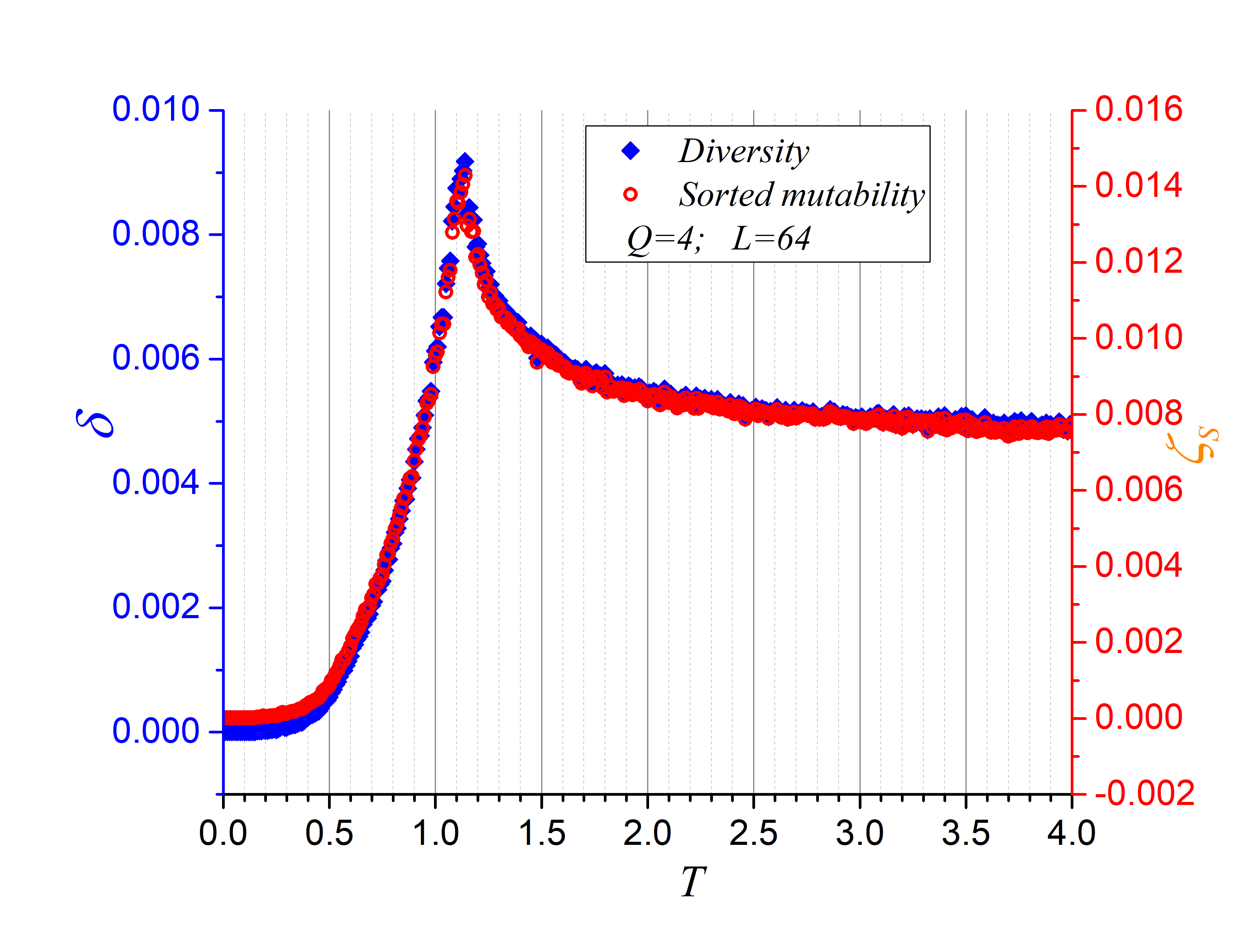}

\includegraphics[width=0.69\columnwidth]{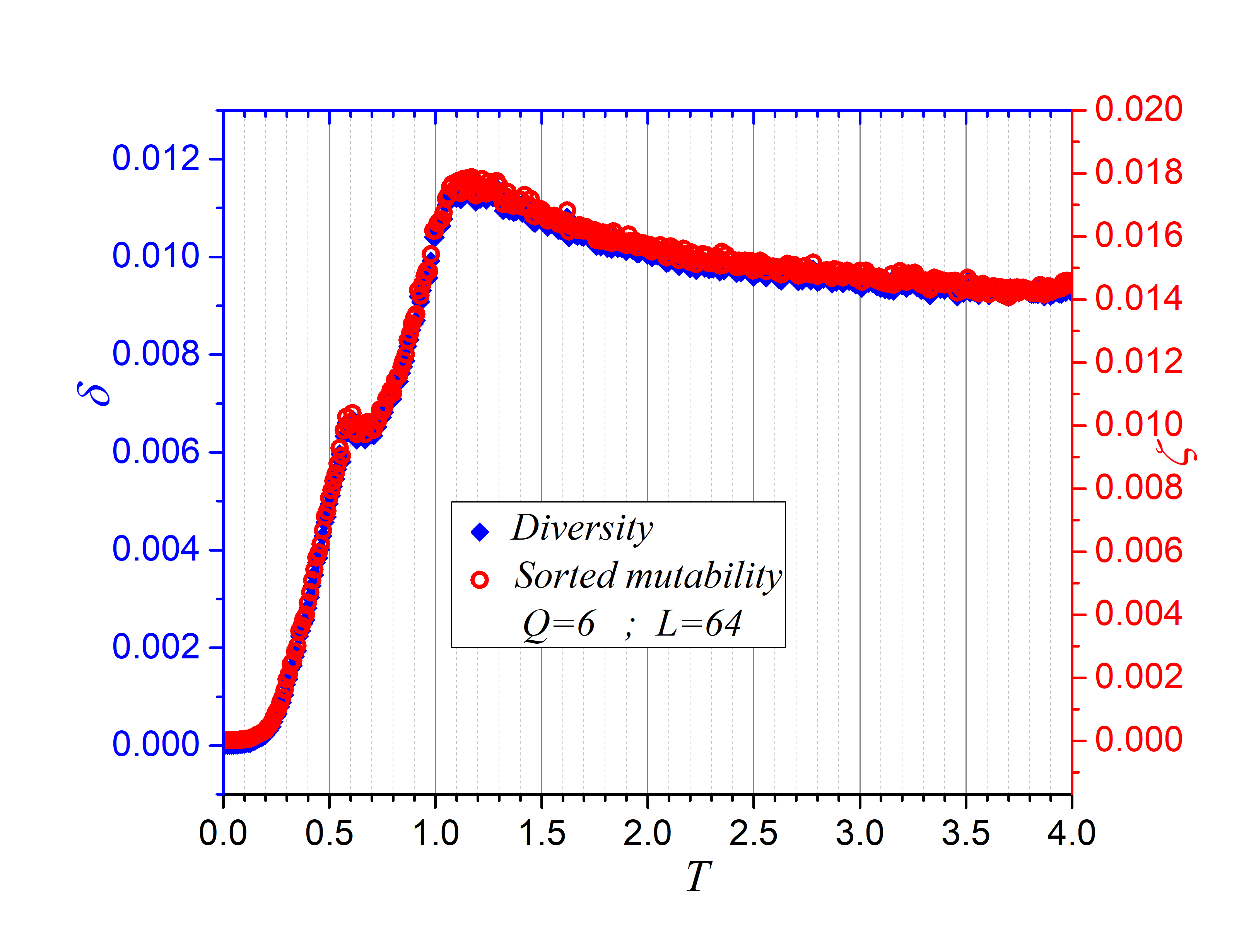}

\caption{Comparison between diversity, $\delta$ (blue diamonds), and sorted mutability, $\zeta_S$ (red circles), as functions of temperature for the $Q$-clock model with $Q=2$, $4$, and $6$, respectively, on square lattices with $L=64$. For every value of $Q$, both quantities exhibit the same temperature dependence up to the representation-dependent scale factor, consistently with $\zeta_S=(b/a)\delta$.}
\label{fig:qclock_Q246}
\end{figure}

Finally, we examine whether the relationship between sorted mutability and diversity persists as the number of allowed spin orientations varies. Figure~\ref{fig:qclock_Q246} compares $\delta(T)$ and $\zeta_S(T)$ for $Q=2$, $4$, and $6$ on square lattices with $L=64$. For every value of $Q$, the two quantities exhibit the same temperature dependence up to the representation-dependent scale factor, consistently with Eq.~(\ref{zeta_s_delta}). An additional calculation for $Q=8$ and $L=128$ is presented in Appendix~B, extending the comparison simultaneously to a larger local configuration space and a larger lattice.

The detailed temperature dependence varies substantially with $Q$, as expected from changes in the accessible local spin states and the corresponding thermal behavior of the clock model. Importantly, however, the proportionality between $\delta(T)$ and $\zeta_S(T)$ remains evident for all values of $Q$ considered. The correspondence is therefore not peculiar to the representative $Q=6$ case but follows from the analytical relation established in Sec.~\ref{sec:theory}.

Taken together, Figs.~\ref{fig:qclock_shannon_sorted}--\ref{fig:qclock_Q246} establish several complementary features of the information-based analysis. First, empirical Shannon entropy and sorted mutability vary over comparable thermal intervals while retaining clearly distinct functional forms, consistent with their probing different properties of the sampled energy distribution. Second, the analytical proportionality between sorted mutability and diversity is recovered numerically through the close correspondence of $\zeta_S(T)$ and $\delta(T)$ and the approximately constant ratio $\zeta_S/\delta$. Third, the principal temperature-dependent structures remain qualitatively stable over the lattice sizes considered. Finally, the correspondence between sorted mutability and diversity persists as the number of allowed clock states is varied.

These observations are deliberately not used to determine phase-transition temperatures from the information-based quantities alone. Instead, the conventional thermodynamic behavior of the $Q$-clock model is presented independently in Appendix~\ref{app:qclock_thermodynamics}, where the energy per spin and specific heat are shown for $Q=2$, $4$, $6$, and $8$ and for lattice sizes $L=64$ and $L=128$. These thermodynamic observables provide an independent reference for determining whether the temperature regions highlighted by the information-based quantities coincide with regions of enhanced energetic or fluctuation-driven activity, without relying on $\zeta_S$, $\delta$, or $H$ themselves as substitutes for finite-size or critical analyses.


\subsection{Wind-power generation}

We consider three time series constructed from the wind-power generation records in Germany: (i) the power generated by onshore wind farms, (ii) the power generated by offshore wind farms, and (iii) the total wind-power generation obtained as the sum of the onshore and offshore contributions. 
Each series contains one record every 15 minutes, corresponding to 96 measurements per day \cite{German_wind}. 
Since a daily sequence would contain only a relatively small number of records for the present information-based analysis, we group the data by month, yielding approximately $3\times10^{3}$ records per monthly sequence, with $\rho$ corresponding to the total number of records used to evaluate $\zeta_S$ and $\delta$ for each month.

The wind-power sequence introduced in Table~\ref{tab:wind} corresponds to a short segment of the total-power series defined in (iii) and was used there only to illustrate the construction of the information-based quantities.

The monthly onshore and offshore power-generation data underlying the present analysis are summarized in Appendix~C. 
Here, we focus on their information-based characterization through sorted mutability and diversity.

\begin{figure}
\centering
\includegraphics[width=0.99\columnwidth]{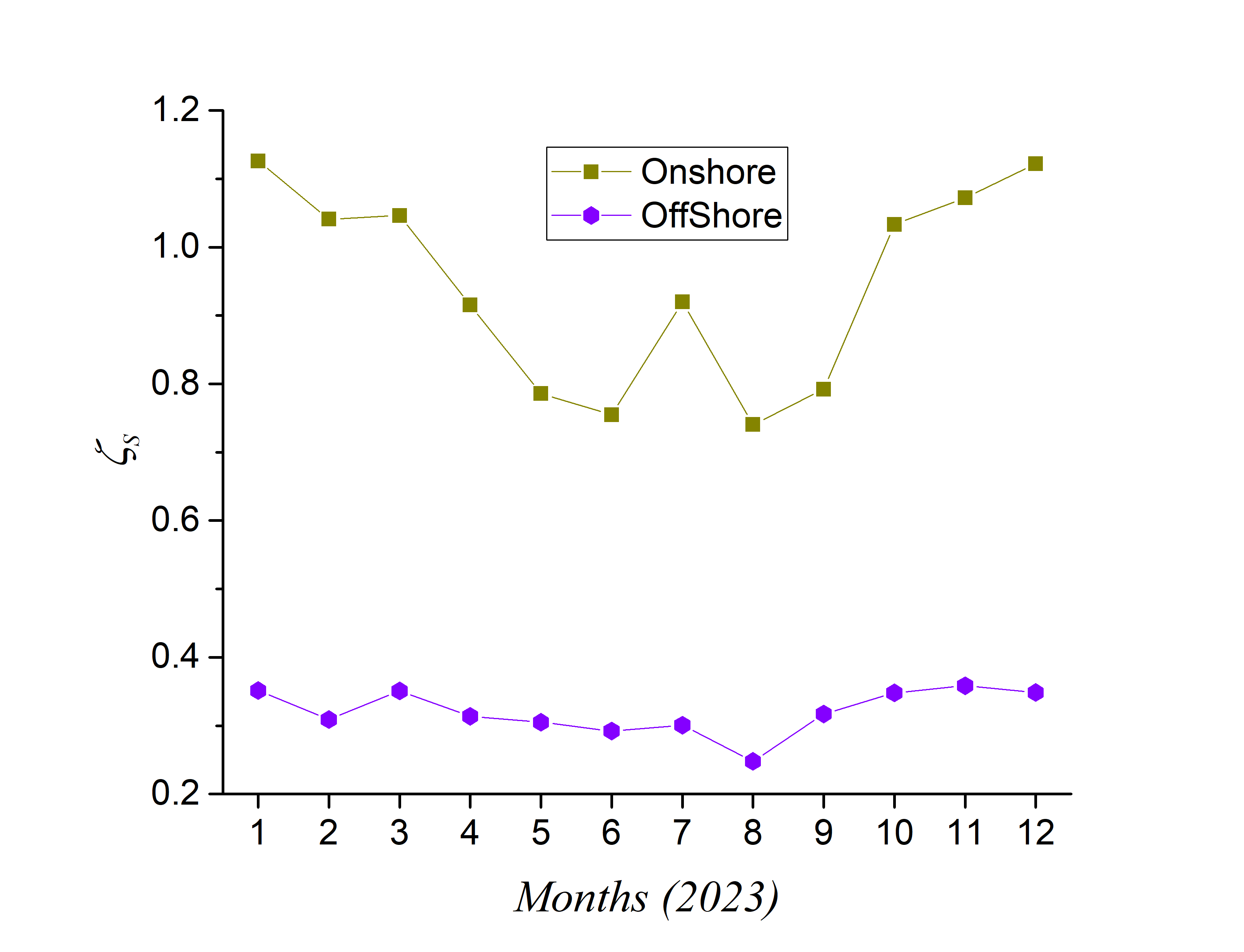}

\includegraphics[width=0.99\columnwidth]{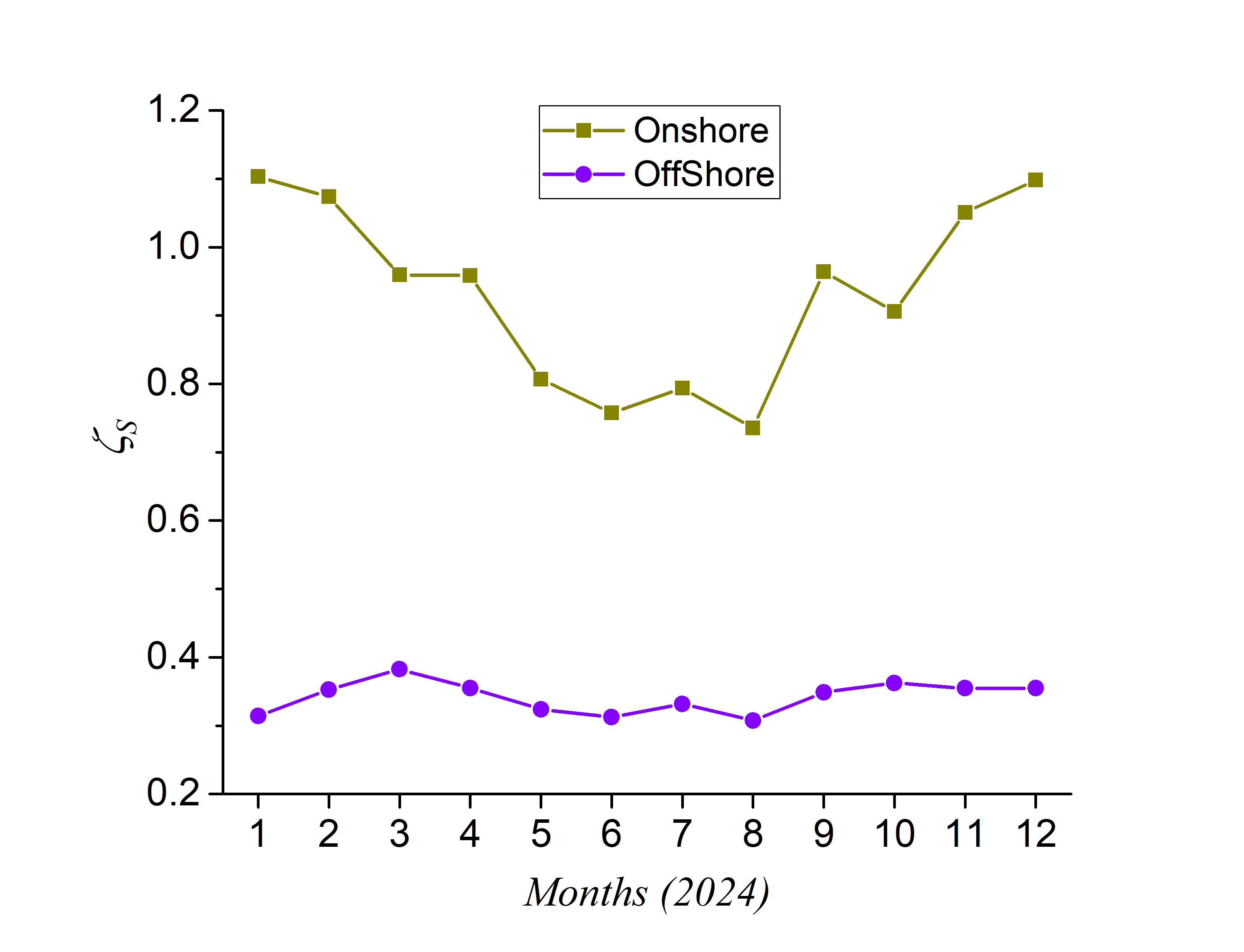}
\caption{
Monthly sorted mutability, $\zeta_S$, of the wind-power generation records in Germany during 2023 (top) and 2024 (bottom). 
Squares and circles correspond to the onshore and offshore contributions, respectively.}
\label{Germany}
\end{figure}

Figure~\ref{Germany} shows the monthly sorted mutability, $\zeta_S$, for the onshore and offshore power-generation series during 2023 and 2024. 
The results for 2023 are displayed in the upper panel, while those for 2024 are shown in the lower panel. 
In both years, the onshore series systematically exhibits larger values of $\zeta_S$ than the offshore series. 
The offshore values remain within a comparatively narrow interval, approximately $\zeta_S\simeq0.30$--$0.35$, whereas the onshore contribution displays a substantially stronger variation throughout the year.

Within the interpretation established in Sec.~\ref{sec:theory}, larger values of $\zeta_S$ indicate a larger fraction of distinct power values represented in the corresponding monthly sequence, for the fixed numerical representation adopted in the analysis. 
Therefore, the difference between the onshore and offshore curves should be understood as a difference in their static diversity rather than as a direct measure of temporal predictability, since the sorting procedure removes the chronological ordering of the records.

The onshore data also display a marked seasonal modulation. 
In both years, relatively large values of $\zeta_S$ occur during several winter months, while lower values are found over much of the summer period. 
Within the present interpretation, this behavior indicates that the winter months tend to explore a comparatively richer set of distinct represented onshore power values. 
Nevertheless, the detailed monthly pattern is not identical from one year to the next. 
For example, differences are visible around midyear and are particularly pronounced in some transition-season months. 
We therefore interpret the seasonal tendency as a qualitative feature of the two-year data set rather than as a strict recurring annual pattern.

\begin{figure}
\centering
\includegraphics[width=0.49\columnwidth]{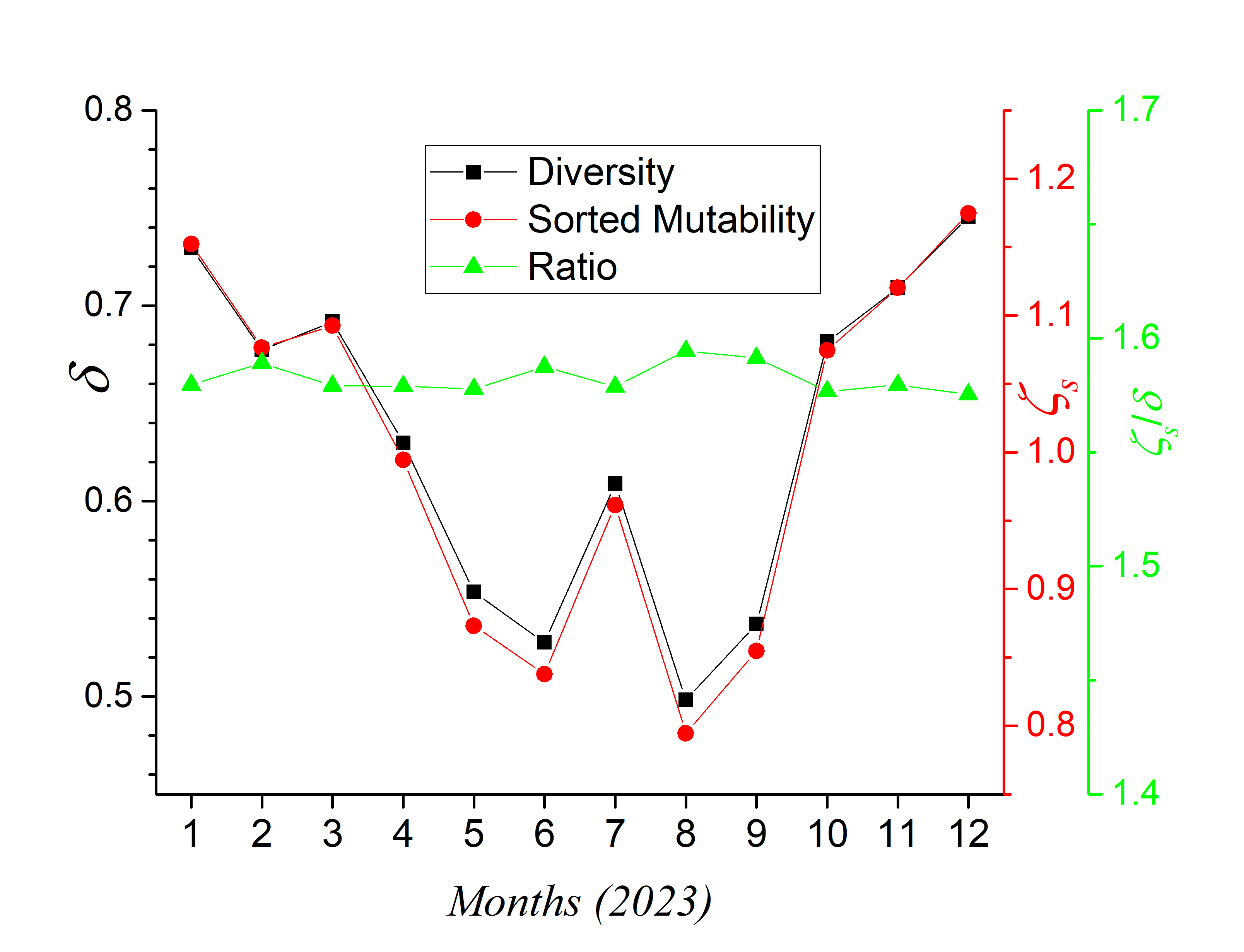}
\includegraphics[width=0.49\columnwidth]{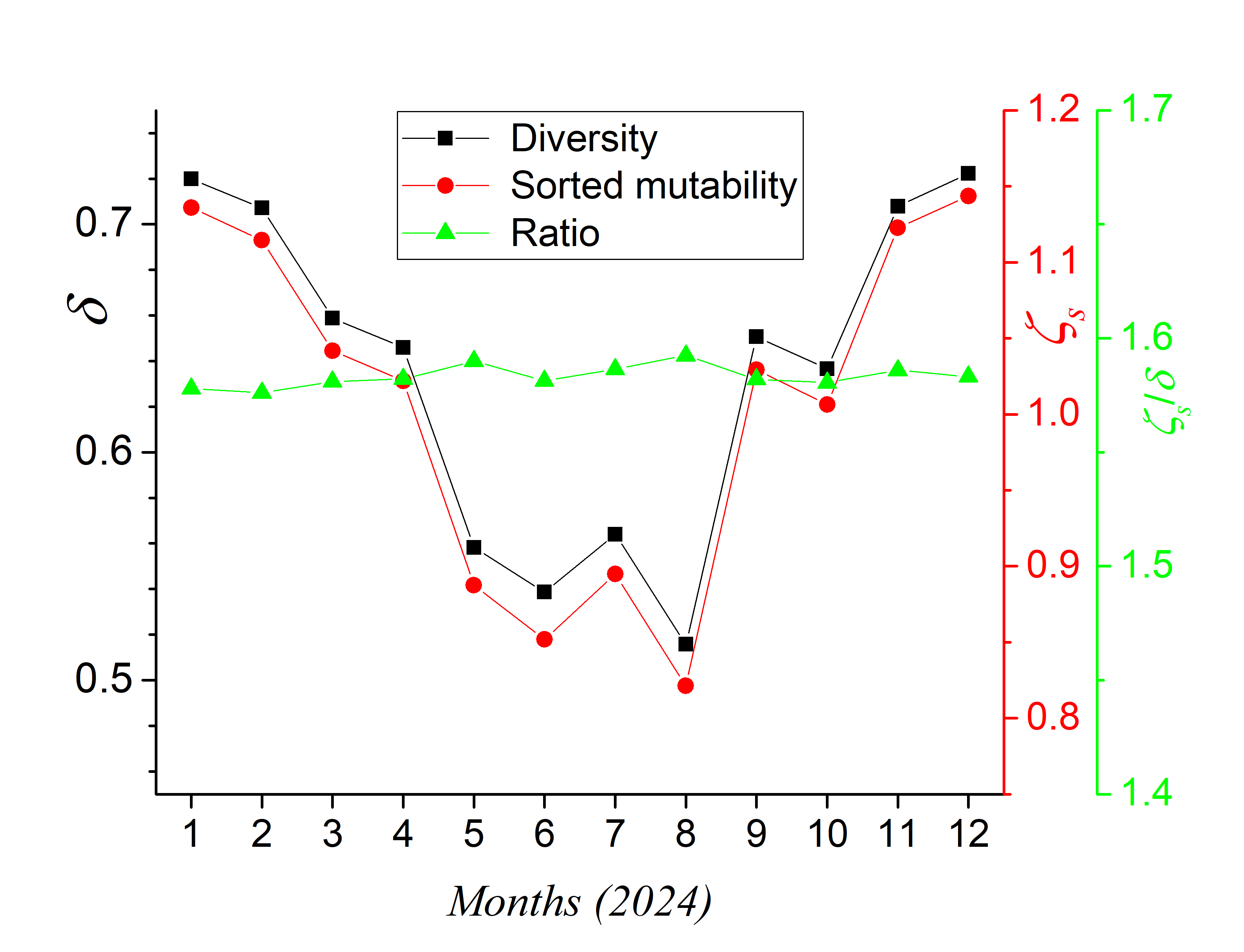}
\caption{
Monthly diversity, $\delta$ (black squares), sorted mutability, $\zeta_S$ (red circles), and their ratio $\zeta_S/\delta$ (green triangles) for the total wind-power generation in Germany during 2023 (left panel) and 2024 (right panel). 
The close correspondence between $\delta$ and $\zeta_S$, together with the nearly constant ratio, illustrates the relation $\zeta_S=(b/a)\delta$ for the adopted numerical representation.
}
\label{zeta_delta_ratio}
\end{figure}

We next examine directly the relation between diversity and sorted mutability for the total wind-power generation. 
Figure~\ref{zeta_delta_ratio} shows the monthly values of $\delta$ and $\zeta_S$ for 2023 (left panel) and 2024 (right panel), together with their ratio $\zeta_S/\delta$. 
The two quantities exhibit the same month-to-month evolution, as expected from Eq.~(\ref{zeta_s_delta}).

The ratio $\zeta_S/\delta$ remains confined to the narrow interval approximately $1.57$--$1.58$ over the two years. 
Thus, as in the $Q$-clock application, the wind-power data provide a numerical consistency check on the proportionality between sorted mutability and diversity, consistent with their definitions. 
The small residual variations observed in $\zeta_S/\delta$ are minor compared with the month-to-month changes of the individual quantities. 
In the present wind-power analysis, each monthly sequence contains approximately $3\times10^{3}$ records, considerably fewer than the $6\times10^{4}$ retained records used in the main $Q$-clock calculations. 
The residual variations of $\zeta_S/\delta$ are therefore more visible in the wind-power data, whereas the substantially longer $Q$-clock sequences yield a more stable numerical ratio.

\section{Conclusions}
\label{sec:conclusions}

In this work, we have established the explicit entropic interpretation of sorted mutability ($\zeta_S$). By demonstrating that, under a fixed numerical representation and encoding prescription, sorted mutability is strictly proportional to the sequence's diversity ($\delta$), we have analytically bridged the gap between compression-based operational metrics and conventional information theory. This mathematical connection allows $\zeta_S$ to be expressed in terms of Shannon entropies. Crucially, because the sorting procedure deliberately destroys temporal correlations, sorted mutability emerges purely as a static information measure that characterizes the richness of the represented states, distinct from temporal organization.

The robustness of this theoretical relation was successfully tested across two fundamentally different physical regimes. In the equilibrium context of the two-dimensional $Q$-clock model, sorted mutability and diversity tracked each other remarkably well across all simulated temperatures, lattice sizes, and allowed spin orientations ($Q=2$ through $Q=8$). While empirical Shannon entropy and sorted mutability respond over comparable thermal intervals, they probe different properties of the energy landscape. More importantly, the temperature regions highlighted by these information-based metrics consistently coincide with critical domains of enhanced energy fluctuations, as independently confirmed by conventional thermodynamic observables such as specific heat. 

The practical utility of this framework is further evidenced by its application to empirical, non-equilibrium data. When analyzing wind power generation in Germany, onshore records exhibited systematically higher and more seasonally modulated sorted mutability than offshore records. Within our established interpretation, this directly quantifies that onshore wind generation explores a significantly richer, more diverse set of static power values. The consistent proportionality between $\delta$ and $\zeta_S$ month-to-month confirms that our theoretical framework translates perfectly to real-world datasets, providing a rapid computational tool for assessing static variability.

Ultimately, sorted mutability provides a robust, highly transferable metric for data diversity across disciplines. Its defining limitation—the deliberate loss of sequential order—is precisely its greatest analytical advantage: it isolates static diversity. This complementary nature suggests a powerful, broader analytical framework for complex systems. By pairing an order-agnostic metric, such as sorted mutability, with an order-sensitive metric, such as regular mutability, future studies in fields ranging from non-equilibrium thermodynamics to econophysics and geophysics can decouple and independently quantify the static richness of a system's states from its underlying temporal dynamics.

\appendix

\section{Thermodynamic reference for the $Q$-clock model}
\label{app:qclock_thermodynamics}

In this Appendix, we provide the conventional thermodynamic behavior of the $Q$-clock systems considered in the main text. The purpose of these results is to establish a reference against which the temperature dependence of the information-based quantities can be compared. Since the model and the Monte Carlo procedure have already been introduced in the main text, we focus here exclusively on the energy and specific heat.

\begin{figure}
    \centering
    \includegraphics[width=0.98\columnwidth]{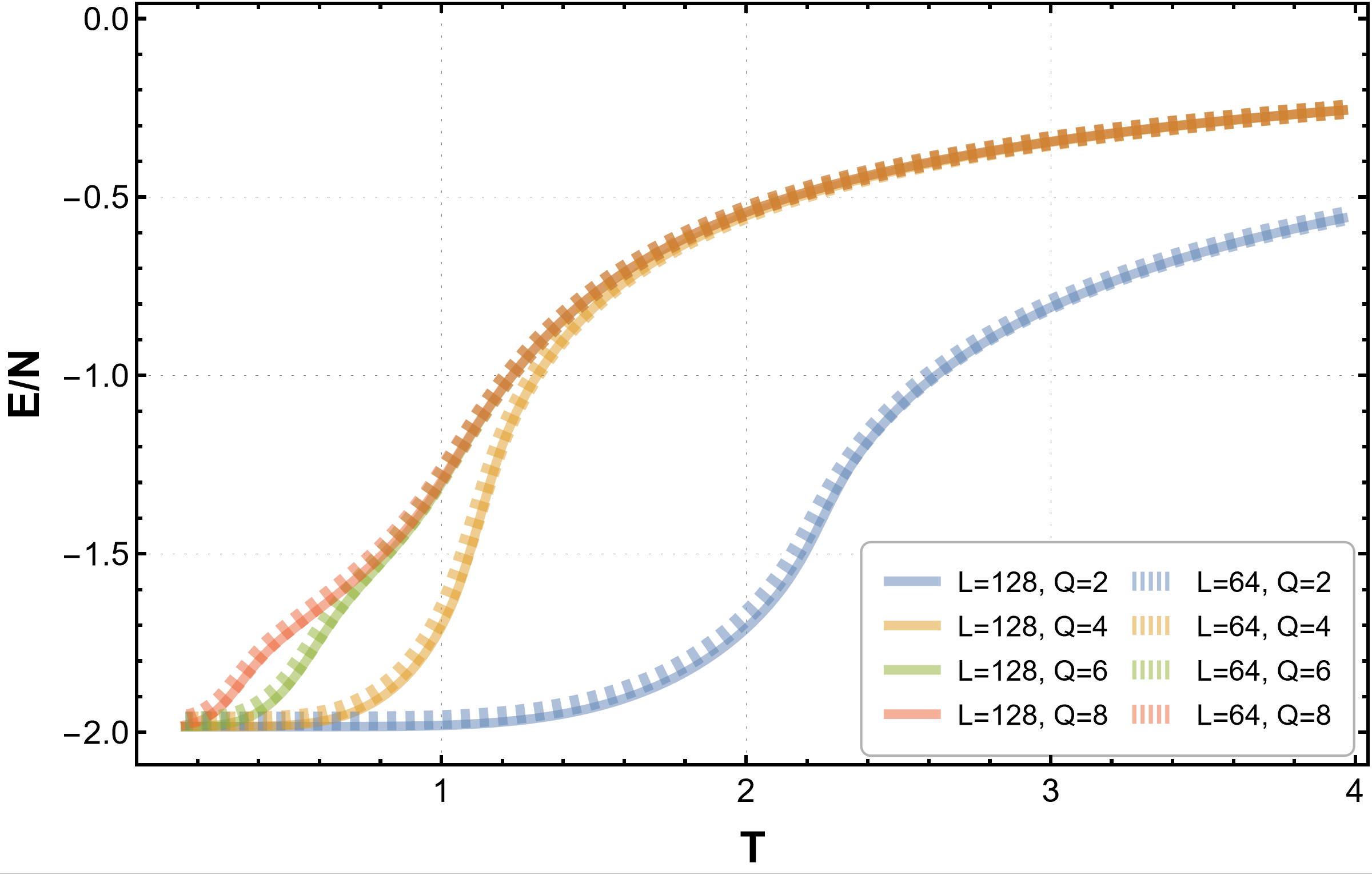}
    \caption{
    Energy per spin, $E/N$, as a function of temperature for the ferromagnetic $Q$-clock model with $Q=2$, $4$, $6$, and $8$. Results for square lattices of linear sizes $L=64$ and $L=128$ are shown for comparison. For each value of $Q$, the curves obtained for the two lattice sizes remain very close over the entire temperature interval, indicating that the overall energetic behavior is already well established for the sizes considered.
    }
    \label{fig:qclock_energy}
\end{figure}

Figure~\ref{fig:qclock_energy} shows the energy per spin, $E/N$, as a function of temperature for $Q=2$, $4$, $6$, and $8$. Results for two lattice sizes, $L=64$ and $L=128$, are included to assess the stability of the thermodynamic behavior with respect to system size.

At low temperature, all systems approach the ordered-state energy $E/N\simeq -2$, as expected for a ferromagnetic square lattice with nearest-neighbor interactions and $J=1$. As the temperature increases, the energy increases as thermally induced spin reorientations progressively reduce the degree of magnetic order. The temperature interval over which this increase becomes most pronounced depends strongly on $Q$.

The $Q=2$ case exhibits the most pronounced energy variation at a substantially higher critical temperature than systems with more accessible spin orientations. For $Q=4$, the rapid change is shifted toward lower temperature. The $Q=6$ and $Q=8$ curves exhibit two critical temperatures, the lower one indicating the ferromagnetic-to-vortex transition and the higher one to the vortex-to-paramagnetic transition. These transitions become more evident in the corresponding specific-heat curves discussed below.

An important observation from Fig.~\ref{fig:qclock_energy} is the close agreement between the $L=64$ and $L=128$ results. Although small finite-size effects may remain near the regions where the energy changes most rapidly, the overall temperature dependence is essentially unchanged upon doubling the linear dimension of the lattice.

\begin{figure}
    \centering
    \includegraphics[width=0.98\columnwidth]{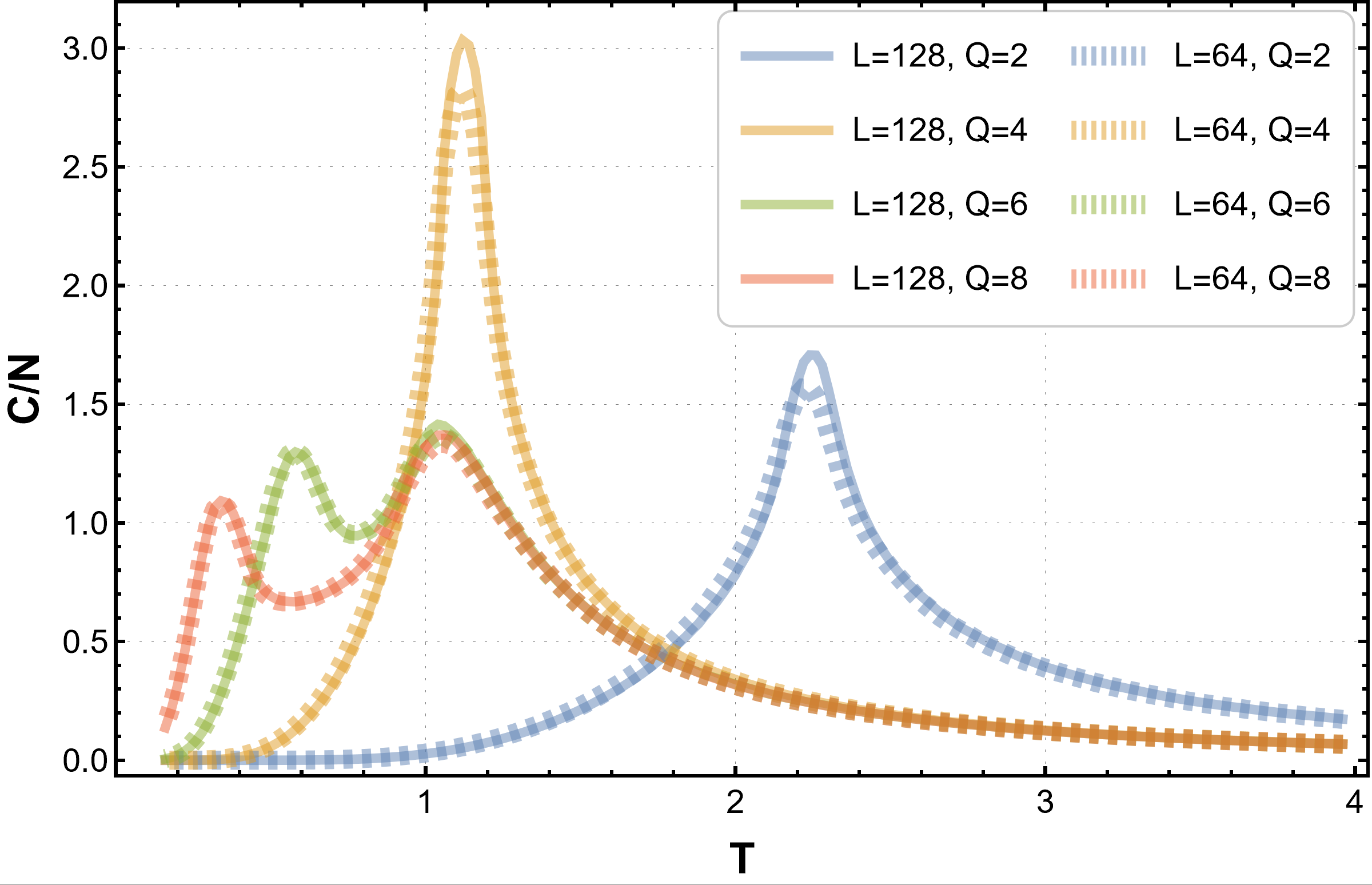}
    \caption{
    Specific heat per spin, $C/N$, as a function of temperature for the ferromagnetic $Q$-clock model with $Q=2$, $4$, $6$, and $8$. Solid and dashed curves correspond to lattice sizes $L=128$ and $L=64$, respectively. The maxima and additional structures identify temperature regions characterized by enhanced energy fluctuations. The close agreement between the two lattice sizes shows that the principal thermodynamic features are robust over the system sizes considered.
    }
    \label{fig:qclock_specific_heat}
\end{figure}

Figure~\ref{fig:qclock_specific_heat} presents the corresponding specific heat per spin, $C/N$, for the same values of $Q$ and the same lattice sizes. In contrast to the relatively smooth behavior of the energy, the specific heat emphasizes the temperature regions where energy fluctuations become particularly important. This is a confirmation of the results obtained for the sorted mutability, i.e $\zeta_S$ as a function of temperature shown in the main text.

For $Q=2$, the specific heat exhibits a single well-defined maximum, coinciding with the temperature interval in which the energy curve develops its steepest variation. A similarly pronounced peak is observed for $Q=4$, although shifted toward lower temperature.

The situation becomes richer for $Q=6$ and $Q=8$. In these cases, the specific heat develops two maxima over the considered temperature range. In particular, a lower-temperature feature corresponding to a particular type of collective transition from the specific-heat data can also be recognized, although less prominently, in the temperature dependence of the energy shown in Fig.~\ref{fig:qclock_energy}.

The comparison between $L=64$ and $L=128$ further shows that the positions and overall shapes of the main specific-heat structures remain stable with increasing lattice size. Finite-size effects are more visible in the amplitudes of the sharper maxima than in their overall location. This behavior supports the use of the $L=64$ simulations in the main information-based analysis, while the $L=128$ results provide an additional check of the robustness of the observed thermal features.

Taken together, Figs.~\ref{fig:qclock_energy} and~\ref{fig:qclock_specific_heat} provide the conventional thermodynamic background for the analysis developed in the main text. They allow the characteristic temperature dependence of Shannon entropy, diversity and sorted mutability to be interpreted relative to the regions in which the underlying $Q$-clock system undergoes its greatest energetic and fluctuation-driven changes. In particular, this separation keeps the central discussion focused on the entropic and information-based quantities while retaining the conventional thermodynamic observables as an independent reference.


\section{Larger lattices}
\label{app:larger_lattices}

In the main text, the comparison between diversity and sorted mutability was presented for $Q=2$, $4$, and $6$ on lattices mainly for the linear size $L=64$. In this Appendix, we include an additional example for a larger lattice and a higher number of allowed spin orientations in order to test the robustness of the same relation under more demanding conditions.

\begin{figure}
\centering
\includegraphics[width=0.69\columnwidth]{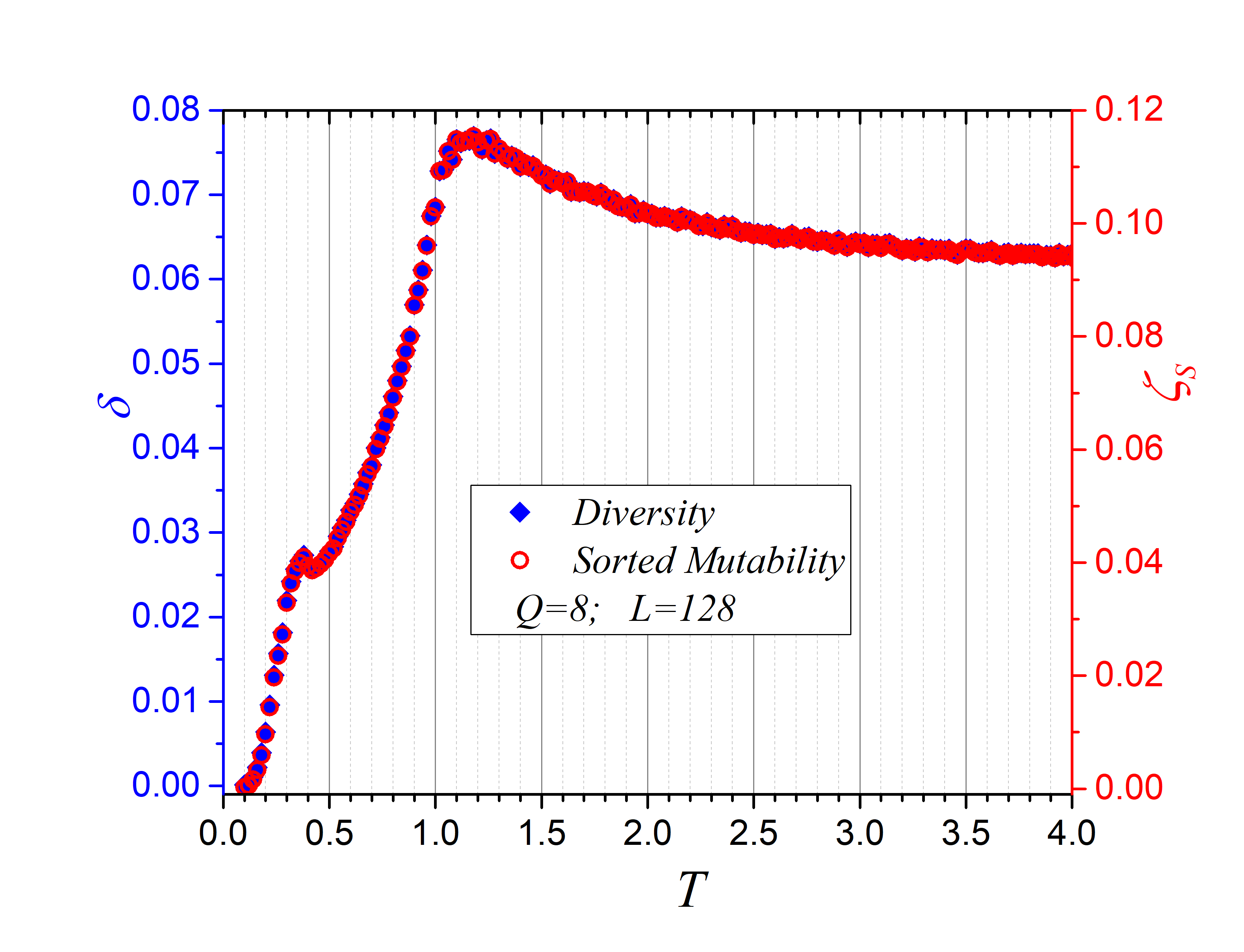}
\caption{
Comparison between diversity, $\delta$ (blue diamonds), and
sorted mutability, $\zeta_S$ (open red circles), as functions of
temperature for the $Q$-clock model with $Q=8$ and lattice size
$L=128$, as indicated in the inset. The close correspondence
between both curves extends the relation $\zeta_S=(b/a)\delta$
to a larger lattice and a higher number of allowed spin
orientations.
}
\label{Q_8_L128}
\end{figure}

Figure~\ref{Q_8_L128} shows the temperature dependence of the diversity, $\delta$, and the sorted mutability, $\zeta_S$, for the $Q$-clock model with $Q=8$ and $L=128$. As in the cases discussed in the main text, both quantities display the same functional dependence on temperature, differing only by an overall scale factor. This behavior is consistent with the proportionality
\begin{equation}
\zeta_S=\frac{b}{a}\,\delta,
\end{equation}
which was established in the theoretical section.

The figure also complements the finite-size discussion of the main text. Together with the results shown previously, the present $Q=8$, $L=128$ case supports the conclusion that the relationship between diversity and sorted mutability is not restricted to small lattices or to a narrow range of $Q$ values. Instead, it remains clearly visible as the system size increases and the local configuration space becomes richer.

Therefore, this additional example reinforces the general character of sorted mutability as an information-based quantity directly linked to diversity, while also illustrating the stability of the results beyond the parameter range emphasized in the main text.


\section{Wind power for 2023 and 2024.}

In the present appendix, we present graphical data on the power series generated in Germany during 2023 and 2024, from which the entropic functions in the main text were obtained.

The data were downloaded from the public database provided by the German system \cite{German_wind}. The information comes every 15 minutes, yielding 96 data points per day and 2880 records over a 30-day month. We prepared three sets of data, separated by calendar month: one with onshore power production, a second with offshore production, and a third with the sum of the first two. 

\begin{figure}
\centering
\includegraphics[width=0.49\columnwidth]{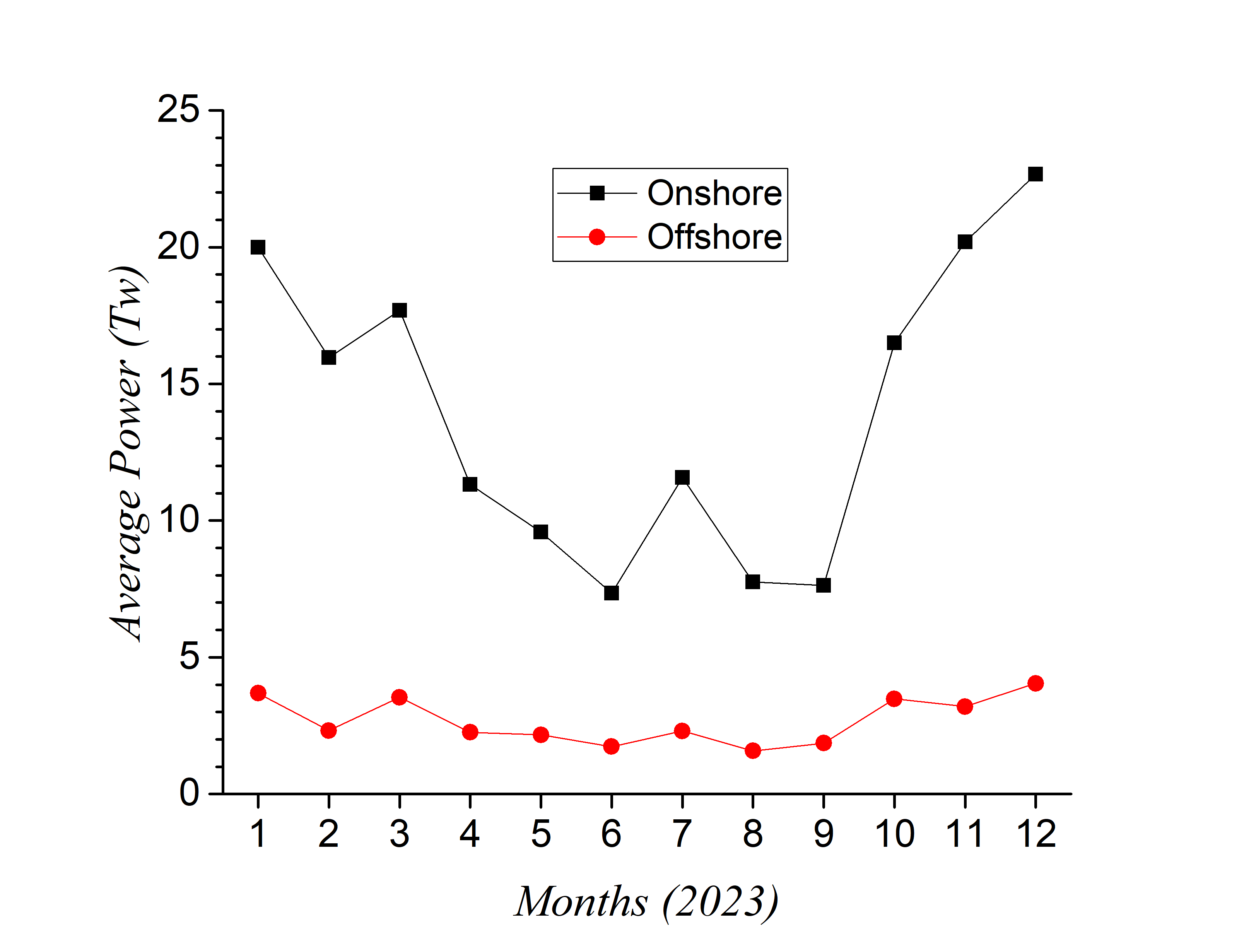} 
\includegraphics[width=0.49\columnwidth]{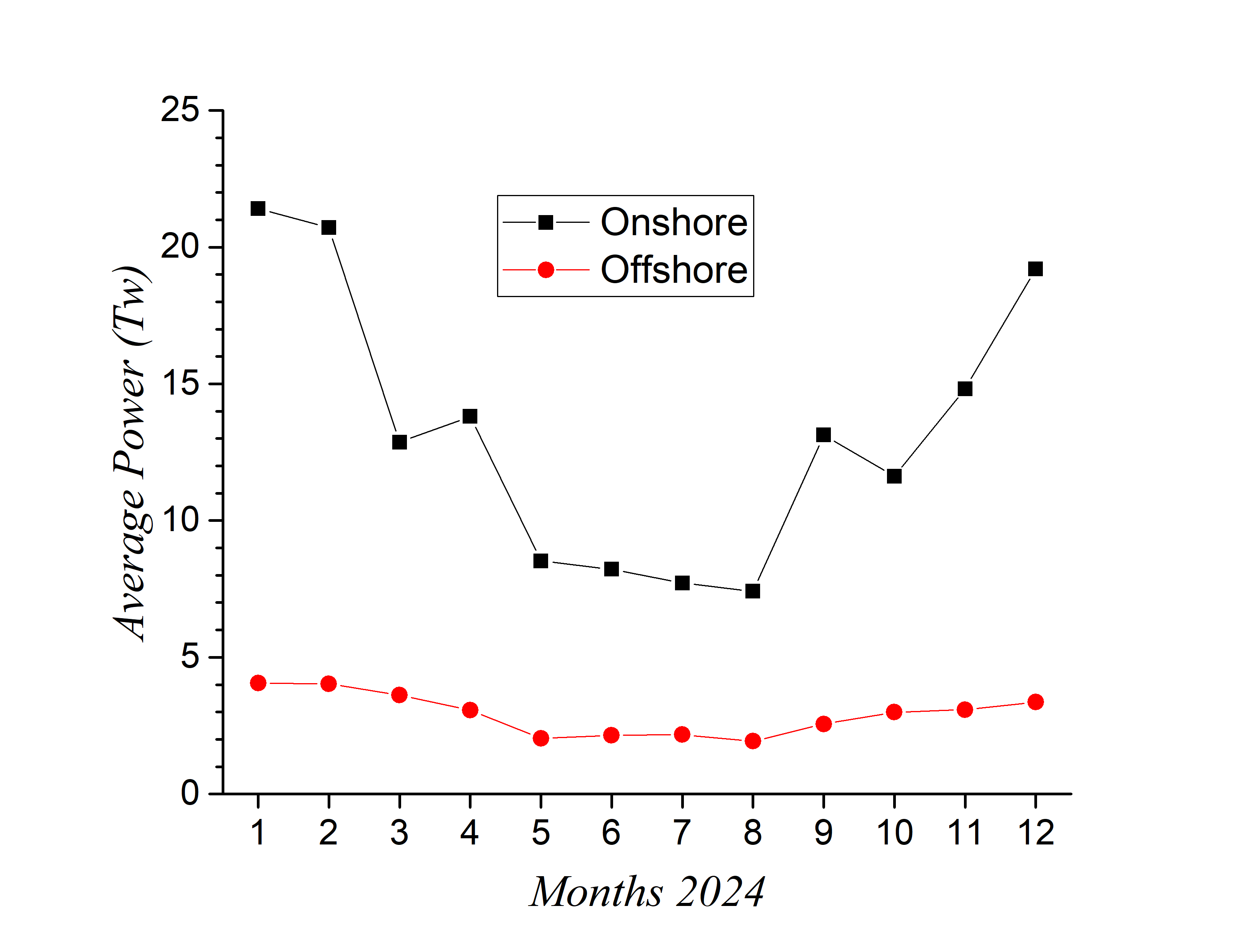} 
  \caption{Monthly average Onshore and Offshore power generation along the months of 2023 (left panel) and 2024 (right panel).
  \label{WEP_2023}}
\end{figure}

Fig. \ref{WEP_2023} presents the average power generated by the German net, separated by the onshore (upper curve) and offshore (lower curve) components; the year 2023 is on the left panel while the year 2024 is on the right panel. We calculated this average per measurement by summing all instantaneous powers and dividing the total by the number of measurements for each month.

The larger winds that occur during winter months make these months more productive, as one would expect. During winter, mutability reaches its maximum, as shown in Fig. \ref{Germany}. This makes wind energy difficult to integrate with other sources of electricity generation: when production is at its peak, its variability is at its highest. Knowledge of how these variations occur can help guide the use of wind power generation.

Offshore wind energy production is more stable, and its variations are also minimal, as presented in Fig. \ref{Germany}. This can make offshore electricity generation easier to combine with other sources. However, logistic considerations make offshore generation less available than onshore generation.

We have used the data originating from wind power generation as an example for the way sorted mutability can be obtained for
any data string. Interested readers on the actual performance of this source of renewable energy are referred to a couple of recent papers and references therein. \cite{vogel2024onshore,Wind_review}

\section*{Acknowledgements} The authors acknowledge the communications and discussions with Prof. Dr. Sigismund Kobe and Mr. Rolf Schuster from Dresden. 
M.A., F.J.P., E.E.V., and P.V. acknowledge the financial support of ANID Fondecyt (Chile) under contracts 1230055, 1240582, and 1250173. M.A. acknowledges the Programa de Incentivo a la Iniciación Científica (PIIC) No. 004/2026 and ANID Becas/Doctorado Nacional 21252820. Authors acknowledge the support from CEDENNA CIA 250002.

--------------

\bibliography{Biblio_entropy}

\end{document}